\documentclass{ws-ijmpcs}
   \usepackage{graphicx}
   \usepackage{amsmath}
   \usepackage{amssymb}
   \usepackage{epstopdf}
   \usepackage{pdfsync}
             \font\sevenrm=cmr7

          \font\sixrm=cmr6

\def\fsc{\alpha_{\hbox{\sevenrm f}}}                                
\def\sigt{\sigma_{\hbox{\sixrm T}}}

\def\dover#1#2{\hbox{${{\displaystyle#1 \vphantom{(} }\over{
   \displaystyle #2 \vphantom{(} }}$}}
\def\teq#1{$\, #1\,$}                         
\def\pmax{p_{\mathrm{max}}} 
\def\gammax{\gamma_{\mathrm{max}}} 
\def\ThetaBfone{\Theta_{\hbox{\sixrm Bf1}}}

\def\betaoneHT{\beta_{\hbox{\sixrm 1HT}}}

\def\betaoneHT{\beta_{\hbox{\sixrm 1HT}}}

\def\today{\ifcase\month\or
  January\or February\or March\or April\or May\or June\or
  July\or August\or September\or October\or November\or
  December\fi
  \space\number\day, \number\year}

\begin{document}

\markboth{M.~G. Baring, M. B\"ottcher \& E. J. Summerlin}
{Multiwavelength Probes of Relativistic Shocks in Blazar Jets}

%
\catchline{}{}{}{}{}
%

\title{MULTIWAVELENGTH PROBES OF THE ENVIRONS 
OF RELATIVISTIC SHOCKS IN BLAZAR JETS}

\author{MATTHEW G. BARING}

\address{Department of Physics and Astronomy - MS 108, Rice University, \\
6100 Main Street, Houston, Texas 77251-1892, USA\\
baring@rice.edu}

\author{MARKUS B\"OTTCHER}

\address{Centre for Space Research, North-West University,\\
Potchefstroom Campus, Potchefstroom, 2520, South Africa\\
Markus.Bottcher@nwu.ac.za}

\author{ERROL J. SUMMERLIN}

\address{Heliospheric Physics Laboratory, Code 672,\\
        NASA's Goddard Space Flight Center, Greenbelt, MD 20770, USA \\  
        errol.summerlin@nasa.gov}

\maketitle

\begin{history}
\received{\today}
\revised{\today}
\end{history}

\begin{abstract}
Diffusive shock acceleration (DSA) at relativistic shocks is likely to
be an important acceleration mechanism in various astrophysical jet
sources, including radio-loud AGN.  An important recent development for
blazar science is the ability of Fermi-LAT data to pin down the
power-law index of the high energy portion of emission in these sources,
and therefore also the index of the underlying non-thermal particle
population.  This diagnostic potential was not possible prior to Fermi
launch, when gamma-ray information was dominated by the highly-absorbed
TeV band. This paper highlights how multiwavelength spectra including
X-ray band and Fermi data can be used to probe diffusive acceleration in
relativistic, oblique, MHD shocks in blazar jets. The spectral index of the
non-thermal particle distributions resulting from Monte Carlo simulations
of DSA, and the fraction of thermal particles accelerated to non-thermal
energies, depend sensitively on the particles' mean free path scale, and
also on the magnetic field obliquity to the shock normal. We investigate
self-consistently the radiative synchrotron/Compton signatures of
the resulting thermal and non-thermal particle distributions.  Important
constraints on the frequency of particle scattering and the level of
field turbulence are identified for the blazar AO 0235+164.  
The possible interpretation that turbulence
levels decline with remoteness from jet shocks, and a significant role
for non-gyroresonant diffusion, are discussed.
\end{abstract}

\ccode{PACS numbers: 52.27.Ny, 52.35.Tc, 52.35.Ra, 97.80.Jp, 95.85.Pw, 98.54.Cm, 98.62.Nx}

\section{Introduction}	
Blazars are a class of active galaxies that are among the most powerful
and dynamic extragalactic objects. They exhibit rapid flares in radio,
optical, X-ray and $\gamma$-ray wavebands.  The general absence of
emission lines in their optical spectra, originally the identifying
feature of Bl Lac objects like {\it BL Lacertae}, together with
detections of polarization in both radio and optical bands, suggests
that the radio-to-optical continuum is non-thermal synchrotron
radiation. In many blazars, this component extends to X-rays. The
watershed development that isolated the class of blazars was the
discovery\cite{Hartmann92,Lin92} of transient gamma-ray emission in 3C
279 and Mrk 421 by the EGRET instrument on the Compton Gamma-Ray
Observatory at around 100 MeV -- 1 GeV. This was almost contemporaneous
with the determination that Mrk 421 also emitted at TeV
energies\cite{Punchetal92}. The prevailing thought is that this
gamma-ray signal is generated by inverse Compton scattering of either an
IR/optical/UV photon source of possibly disk or ambient
origin\cite{DSM92,SBR94}, or of synchrotron photons by the same
electrons that emit this radio-to-X-ray signal, so-called
synchrotron-self-Compton (SSC) models\cite{MCG92,MK97,CB02}. An
important recent development for blazar science has been the improvement
of sensivity in the 100 MeV-100 GeV window, afforded by the {\it
Fermi}-Large Area Telescope (LAT).  Over the last five years, LAT data
has enabled measurements of the power-law index of blazar
spectra\cite{Abdo09,Abdo10}, and therefore also the index of the
underlying non-thermal particle population.

The flux variability and enormous power of blazars indicate that they
possess central supermassive black holes with masses typically above
\teq{10^8M_{\odot}}, and are characterized by powerful jets that emanate
from near the event horizon and point almost directly towards Earth. The
rapid variability seen in GeV--TeV flares drives the prevailing picture
for the blazar environment: their jets are relativistic, and compactly
structured on small spatial scales that are unresolvable by present
gamma-ray telescopes. Velocity structure in the supersonic outflow in
these jets naturally  generates relativistic shocks, and these can form
the principal sites for acceleration  of electrons and perhaps ions to
the ultrarelativistic energies demanded by the X-ray and $\gamma$-ray
data\cite{BE87}. Diffusive acceleration at such shocks is believed to be
the main mechanism for energizing such charges\cite{BE87,Drury83,JE91}.
This is because it is both extremely efficient and very fast,
precipitating acceleration rates \teq{{\dot \gamma}} of the order of the
gyrofrequency \teq{\omega_g = eB/mc}, since the diffusive collisions of
electrons and ions with MHD turbulence in jets is usually dominated by
gyroresonant interactions. The diffusive mean free path \teq{\lambda} of
a charge is then at least comparable to its Larmor radius \teq{r_g =
pc/eB =\gamma\beta mc^2/eB} if it has a momentum \teq{p=\gamma\beta mc}
and Lorentz factor \teq{\gamma = 1/\sqrt{1-\beta^2}}; the \teq{\lambda
\sim r_g} regime is known as the Bohm limit. This scenario invokes a
conversion of the ballistic kinetic energies of colliding, large-scale
MHD structures in jets into the energization of \teq{\gamma\gg 1}
particles at relativistic shocks. Other paradigms such as acceleration
by reconnection of magnetic fields embedded in Poynting-flux dominated
outflows can be envisaged. Yet, there is presently no compelling reason
to disavow the shock acceleration picture for blazar jets.

To explore the shock paradigm, the standard practice has been to develop
multiwavelength (MW) spectral models, spanning radio to TeV gamma-ray
wavelengths\cite{DSM92,SBR94,MCG92,MK97,CB02,BDF08}. This generally
gives a broad-brush assessment of a range of jet environment parameters,
but the spectral fits in any one band are of limited quality. The high
statistics spectroscopy offered by the {\it Fermi-LAT} data demands a
closer look at the constraints observations can make on the shock
acceleration process and the blazar jet environment.  For example, it
has long been known that non-relativistic shocks generate a power-law
index that is dependent only on the hydrodynamic velocity compression
ratio \teq{r} across the shock\cite{Drury83,JE91}. For strong shocks
with \teq{r=4}, this yields a \teq{p^{-2}} distribution. This robustness
has helped underpin a consensus that Galactic cosmic rays originate in
supernova remnants. In contrast, the convective loss rates downstream of
relativistic shocks are sensitive to the orientation of the mean
magnetic field, and the character of the {\it in situ} MHD turbulence,
yielding a wide variety of power-law indices for accelerated charges for
a given velocity compression ratio\cite{EJR90,KH89,ED04,SB12}.
Accordingly, gamma-ray spectroscopy of blazars can afford insights into
details of the structure of shocks in jets.  In addition, it has been
understood for nearly two decades\cite{IT96} that diffusive shock
acceleration is so efficient that low levels of field turbulence
(spawning \teq{\lambda\gg r_g}) are required to accommodate synchrotron
spectral peaks appearing in the X-ray band.

In gleaning such insights, multi-wavelength blazar models generally
invoke an elemental description of the distribution of accelerated
charges, usually a power-law truncated at some minimum particle Lorentz
factor.  Such forms omit the hot Maxwellian-like components that
naturally arise due to shock layer turbulent dissipation of the bulk
kinetic energy of the upstream flow.  This paper goes beyond such
simplified approaches by employing detailed results from comprehensive
Monte Carlo simulations of diffusive acceleration at relativistic
shocks, building upon the exposition of Summerlin \& Baring\cite{SB12}. 
These simulations capture the relationship between turbulence parameters
and the power-law index, and also the connection between the thermal
bulk of the population and the power-law tail of the accelerated
species. We fold these results through the one-zone SSC/external Compton
models of B\"ottcher et al.\cite{Boettcher13} for radiation emission and
transport in blazars. In particular, diagnostics are obtained for the
frequency of particle scattering and the level of field turbulence for
the Bl Lac object AO 0235+164. By exploring a range of dependences of
the diffusive mean free path \teq{\lambda} on the momentum \teq{p} of
accelerated charges, the possible interpretation that turbulence levels
decline with remoteness from a shock is identified, perhaps signalling a
significant role for non-gyroresonant diffusion near blazar jet shocks.

\newpage

\section{Shock Acceleration Simulations of Particle Distributions}

In collisionless shocks, non-thermal, charged particles gain energy by
scattering between MHD turbulence that is ``anchored'' in the converging
upstream and downstream plasmas. In non-relativistic shocks, the
energetic particles are nearly isotropic in these fluid frames, since
their speeds \teq{v\approx c} far exceed \teq{u_{1x}} (\teq{u_{2x}}),
the upstream (downstream) flow speed component in the co-ordinate
direction \teq{x} normal to the shock.  The acceleration process
possesses no momentum scale, and the resulting distribution takes the
form \teq{dn/dp \propto p^{-\sigma}}, where \teq{\sigma = (r+2)/(r-1)},
for \teq{r=u_{1x}/u_{2x}} being the shock's velocity compression ratio. 
The index \teq{\sigma} in this limit is independent of the shock speed,
\teq{u_1}, the upstream field obliquity angle \teq{\ThetaBfone} (to the
shock normal), and any details of the scattering process. In this
test-particle limit, high Mach number non-relativistic shocks have
\teq{r \simeq 4} yielding \teq{dn/dp \propto p^{-2}}.

In contrast, in relativistic shocks, plasma anisotropy is prevalent, and
\teq{\sigma} becomes a function of the flow speed \teq{u_1}, the field
obliquity, and the nature of the scattering. Test-particle acceleration
in parallel relativistic shocks evinces the key property that a
so-called ``universal'' spectral index\cite{BO98,Baring99,Kirk00},
\teq{\sigma\sim 2.23} exists in the two limits of \teq{\Gamma_1 \gg 1}
and small angle scattering, i.e., \teq{\delta \theta \ll 1/\Gamma_1},
for a shock compression ratio of \teq{r=3}. Here \teq{\delta\theta} is
the average angle a particle's momentum vector deviates in a scattering
event, and $\Gamma_1 = (1 - [u_1/c]^2)^{-1/2}$ is the Lorentz factor of
the upstream flow in the shock rest frame.  For all other parameter
regimes in relativistic shocks, large departures from this special index
are observed.  Note that relativistic shocks in blazar environs are
typically highly oblique (\teq{\ThetaBfone \gtrsim 1/\Gamma_1}) due to
the Lorentz transformation of ambient, upstream magnetic fields to the
shock rest frame.

Various approaches have been adopted to model diffusive shock
acceleration (DSA) at relativistic discontinuities. These include
analytic methods\cite{Peacock81,KS87,Kirk00}, and Monte Carlo
simulations of convection and diffusion\cite{EJR90,BO98,ED04,NO04,SB12}.
Particle-in-cell (PIC) plasma simulations have also been employed to
study this problem via modeling the establishment of turbulence
self-consistently\cite{Gallant92,Nishikawa03,SS09}, though they cannot
yet explore acceleration beyond modest non-thermal energies. The Monte
Carlo approach is ideal for describing the diffusive acceleration of
particles from thermal (injection) scales out to the highest energies
relevant to blazar jet models.

The Monte Carlo technique employed here solves the Boltzmann equation
with a phenomenological scattering operator\cite{JE91,EJR90,SB12}.  The
simulation space is divided into grid sections with different flow
speeds, magnetic fields, etc. For electron-ion shocks, the vastly
different inertial scales can be easily accommodated. Particles are
injected far upstream and diffuse and convect in the shock neighborhood.
 Their flux-weighted contributions to the momentum distribution function
are logged at any position \teq{x}. Each particle is followed until it
leaves the system by either convecting far downstream, or exceeding some
prescribed maximum momentum \teq{\pmax}.

The details of particle transport are given in\cite{EJR90,ED04,SB12}:
the simulation can model both scatterings with small angular deflections
\teq{\delta\theta \ll 1/\Gamma_1} (seeding pitch angle diffusion) and
large ones (\teq{\delta\theta \gtrsim 1/\Gamma_1}) using several
parameters. In this paper, the focus will be on \teq{\delta\theta \ll
1/\Gamma_1} domains; results for large angle scattering are illustrated
in\cite{EJR90,SBS07}. The simulation assumes that the complicated plasma
physics of wave-particle interactions can be described by a simple
scattering relation for individual particles, viz.
\begin{equation}
   \lambda_{\parallel} \; =\; \lambda_1 \, \dover{\rho_1}{\rho} \left ( \dover{r_g}{r_{g1}} \right )^\alpha
   \;\equiv\; \eta_1 \left( \dover{p}{p_1} \right)^{\alpha}
   \quad \hbox{{\rm and}}  \quad
   \kappa_{\parallel} = \lambda_{\parallel} v/3\ ,
 \label{eq:mfp_alpha}
\end{equation}
where \teq{v=p/m} is the particle speed in the local upstream or
downstream fluid frame, \teq{r_g =pc/(QeB)} is the gyroradius of a
particle of charge $Qe$, and $\rho$ is the plasma density, with a far
upstream value of \teq{\rho_1}. Here, \teq{\lambda_\parallel}
(\teq{\kappa_{\parallel}}) is the mean free path (spatial diffusion
coefficient) in the local fluid frame, parallel to the field {\bf B},
with \teq{\lambda_\parallel\gtrsim r_g} being a fundamental bound (Bohm
limit) for physically meaningful diffusion.  Scalings of the momentum
\teq{p_1=mu_{1x}=m\beta_{1x}c} and the mean free path \teq{\lambda_1=
\eta_1 r_{g1}} for \teq{r_{g1} =p_1c/(QeB_1)} are introduced to simplify
the algebra.

Scattering according to Eq.~(\ref{eq:mfp_alpha}) is equivalent to a
kinetic theory description\cite{FJO74} where the diffusion coefficients
perpendicular to (\teq{\kappa_{\perp}}) and parallel to
(\teq{\kappa_{\parallel}}) {\bf B} are related via \teq{\kappa_{\perp} =
\kappa_{\parallel} / [1 + (\lambda_{\parallel}/r_g)^2]}. The parameter
\teq{\eta \equiv \lambda_{\parallel}/r_g} then characterizes the
``strength" of the scattering and the importance of cross-field
diffusion:  when \teq{\eta \sim 1} at the Bohm Limit,
\teq{\kappa_{\perp} \sim \kappa_{\parallel}} and particles diffuse
across magnetic field lines nearly as quickly as along them. Each
scattering event is an elastic deflection of the fluid frame momentum
vector {\bf p} through angle \teq{\sim \delta\theta}, so that the number
of deflections constituting \teq{\lambda_\parallel}, a large angle
deflection scale, is proportional to \teq{(\delta\theta)^{-2}}.

Observations at the Earth's bow shock\cite{EMP90} indicate that \teq{1/2
< \alpha < 3/2}, while \teq{\alpha\sim 1/2} is obtained from turbulence
in the interplanetary magnetic field\cite{Moussas92}.  Hybrid plasma
simulations\cite{GBS92} suggest a mean free path obeying
Eq.~(\ref{eq:mfp_alpha}) with \teq{\alpha \sim 2/3}. Within the confines
of quasi-linear theory of MHD turbulence, diffusion characteristics
embodied in Eq.~(\ref{eq:mfp_alpha}) can be coupled\cite{FJO74} to the
inertial range and power spectrum \teq{\langle \delta B/B\rangle^2} of
wave numbers {\bf k} for the turbulence.  These solar wind observations
and plasma simulation results explore domains where the scales of
turbulence {\bf k} and diffusion \teq{\lambda_{\parallel}} are very
restricted.  The story that will unfold in this paper is that blazar
jets may present a very different picture with their associated large
dynamic ranges of momenta and mean free paths in relativistic jets.

\begin{figure}[pt]
\centerline{\includegraphics[width=11.0cm]{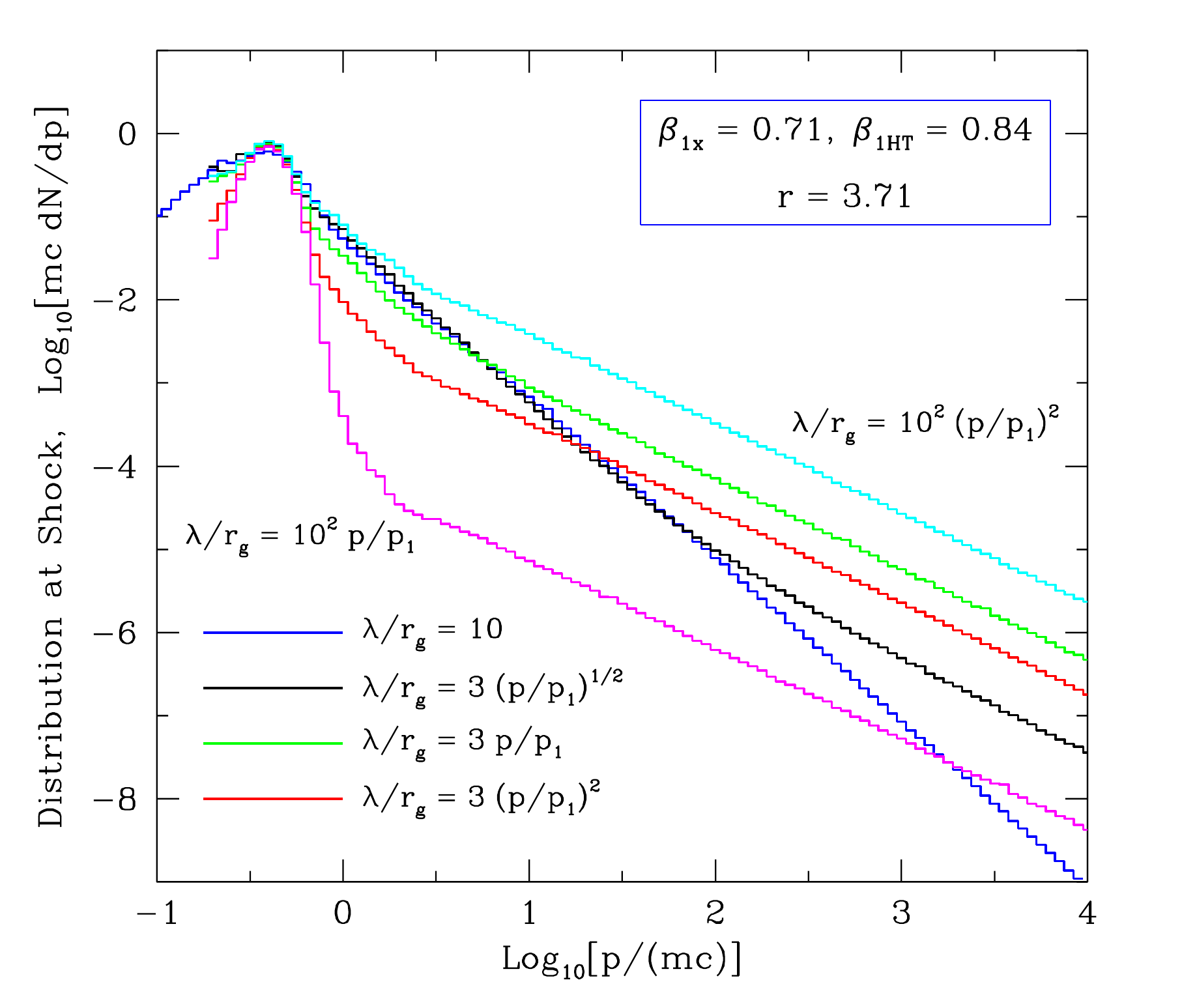}}
\vspace*{-5pt}
\caption{Particle distributions for acceleration simulation runs in the 
small angle scattering limit, for strong mildly-relativistic shocks of
upstream flow speed \teq{\beta_{1x}\equiv u_{1x}/c =0.71}.  Here 
the de Hoffmann-Teller frame upstream flow speed was set at
\teq{\betaoneHT =0.75=\beta_{1x}/\cos\ThetaBfone}, with 
\teq{\ThetaBfone \approx 48.2^{\circ}} being the upstream field 
obliquity to the shock normal.  Distributions are displayed for six 
different forms for the momentum dependence of the diffusive 
mean fee path \teq{\lambda\equiv\lambda_{\parallel}}, namely 
\teq{\lambda/r_g \propto p^{\alpha -1}} with \teq{\alpha -1 = 0, 1/2, 1} and \teq{2},
as labelled --- see Eq.~(\ref{eq:mfp_alpha}) and associated discussion.  The shock
velocity compression ratio was fixed at \teq{r=u_{1x}/u_{2x}=3.71}, and the 
upstream temperature corresponded to a sonic Mach number of \teq{{M}_{\rm S}\sim 4}.  
See Fig.~10 of Ref. [18] for more \teq{\alpha =1} cases.
\label{fig:accel_dist}}
\end{figure}

In relativistic shocks, the distribution functions of accelerated
charges are sensitive to the choices of both the
\teq{\eta_1=\lambda_1/r_{g1}} and \teq{\alpha} diffusion parameters. 
The \teq{\eta} dependence of acceleration is illustrated at length in
Summerlin \& Baring\cite{SB12}, where \teq{\alpha = 1} is a restriction
imposed for simplicity, so that \teq{\lambda_{\parallel} \propto r_g}. 
In subluminal shocks, those where a de Hoffman-Teller (HT) frame\cite{HT50} 
can be found (i.e., \teq{u_{1x}/\cos\ThetaBfone <c},
where again, \teq{\ThetaBfone} is the upstream field obliquity angle to
the shock normal), distributions \teq{dn/dp \propto p^{-\sigma}}
generally possess indices in the range \teq{1 < \sigma < 2.5}.  The HT
frame is where the shock is at rest and the fluid flows along {\bf B}.
An example of such a spectrum is the \teq{\eta=10} case in
Fig.~\ref{fig:accel_dist}, where the power-law tail smoothly blends into
the upper end of the thermal distribution.  For much of the fairly
restricted range of obliquities corresponding to
\teq{u_1/\cos\ThetaBfone <c}, the larger the value of \teq{\eta}, the
lower \teq{\sigma} is.  In fact, when \teq{\eta\gg 1}, \teq{\sigma\sim
1} is generally observed\cite{SB12} and the acceleration distribution is
extremely flat.  The origin of this behavior was found to be shock drift
acceleration, where charges with select gyrational phases incident upon
the shock from upstream, are trapped and reflected by the shock.  Shock
drift acceleration is a phenomenon that has been well-studied in the
context of heliospheric shocks\cite{Jokipii82,DV86}. Successive episodes
of upstream excursions and reflection at the shock precipitate efficient
acceleration and postpone convective loss downstream for many shock
interaction cycles. Only when the field obliquity \teq{\ThetaBfone} is
high enough that the shock is superluminal and \teq{u_1/\cos\ThetaBfone
>c} does the powerful convective action of the flow overwhelm
reflection, and the acceleration process effectively shuts
off\cite{ED04,SB12,BK90}, with the index \teq{\sigma} increasing rapidly
above \teq{3-4}.

Here, these results are extended to situations where \teq{\eta (p)} is
an increasing function of momentum \teq{p}.  The simplest prescription
that captures such character is Eq.~(\ref{eq:mfp_alpha}), and
representative \teq{\alpha >1} examples are displayed in
Fig.~\ref{fig:accel_dist}.  The subluminal shock parameters used therein
are mostly those in Fig.~10, left panel, of Ref. [\refcite{SB12}].  The
distributions all display the generic trait of a dominant thermal
population with a power-law tail that extends as high as the geometric
scale of the diffusive acceleration zone permits: this environmental
parameter is addressed in Section~\ref{sec:radmodels}. Each distribution
possesses an injection efficiency from thermal into the Fermi process
that reflects the value of \teq{\eta_1} in Eq.~(\ref{eq:mfp_alpha}).
Large values of \teq{\eta_1} produce a low normalization to the
power-law at supra-thermal momenta \teq{p\sim 3p_1-10p_1}.  The ultimate
power-law index at high momenta \teq{p\gg p_1} is determined by large
values for \teq{\eta (p)} in all these simulation runs, i.e., realizing
\teq{\sigma\sim 1} regimes in weak turbulence in
Fig.~\ref{fig:accel_dist} due to efficient shock drift acceleration. 
The key new signature presented here is the appearance of ``flattening''
breaks in the high energy tail that are manifested when \teq{\eta (p)}
begins to exceed values around \teq{10-30}; see the \teq{\eta_1=3},
\teq{\alpha =3/2} and \teq{\eta_1=3}, \teq{\alpha =2} cases in
Fig.~\ref{fig:accel_dist}. Hence, \teq{\alpha >1} models can possess
relatively inefficient injection at thermal energies, with very dominant
thermal components, but exhibit efficient acceleration at momenta
\teq{p\gtrsim 100p_1}. The properties of plasma turbulence that might
generate \teq{\alpha >1} circumstances are discussed below in
Section~\ref{sec:discussion}.

\section{Radiation Emission Modeling}
 \label{sec:radmodels}

The focus now turns to modeling the broadband continuum emission of the
blazar AO 0235+164 using specific distributions generated by the Monte
Carlo technique.  The objective is to assess how the global character of
the radio to X-ray to gamma-ray data can provide insights into the
nature of the accelerator and its plasma environment.  To facilitate
this goal, we focus solely on leptonic
models\cite{MCG92,MK97,CB02,Boettcher13}, those with synchrotron
emission predominantly at frequencies below \teq{10^{19}}Hz, and with
inverse Compton emission dominating the gamma-ray signal. Hadronic
models are also of great interest, but then introduce more parameters,
and so don't tend to provide constraints that are as restrictive as
those explored here.

To evaluate the emission from the Monte Carlo generated electron
distributions described in the previous section, we adopt radiation
modules from the one zone blazar radiation transfer code of B\"ottcher
et al.\cite{Boettcher13,BC02}. This code treats synchrotron emission,
and also inverse Compton scattering of both synchrotron and external
seed photons. Bremsstrahlung is modeled in the code, but is usually
insignificant in blazars: it is so for AO 0235+164. For this purpose, we
first note that without the introduction of information on the physical
scale of the acceleration zone, the Monte Carlo simulations do not
describe the high-energy cut-off of the electron distribution. We
therefore extend the initial electron spectra like those in
Fig.~\ref{fig:accel_dist} out to a maximum energy \teq{\gamma_{\rm
max}mc^2 \approx \pmax c}.  Note that such an extrapolation is
acceptable because the asymptotic power-law has been realized in the
Monte Carlo simulations. This maximum energy (a model parameter) is
constrained in two ways: (a) particles will not be accelerated beyond an
energy for which the (synchrotron plus inverse Compton) radiative
cooling time scale, \teq{t_{\rm rad} = 3 m_e c^2 / (4c \, \sigt \, {\cal
U} \, \gamma)}, where \teq{{\cal U} = {\cal U}_B + {\cal U}_{\rm rad}}
is the sum of the energy densities in the magnetic field and the photon
field, is shorter than the acceleration time scale, \teq{t_{\rm acc} =
\gamma m_e c \, \eta( p) / (e B)}; and (b) particles will not be
accelerated to energies at which the diffusive mean free path
\teq{\lambda_{\parallel} = \eta(p) r_g(\gamma)\equiv \eta_1\,
(p/p_1)^{\alpha}} in Eq.~(\ref{eq:mfp_alpha}) exceeds the size
\teq{R_{\rm acc}} of the acceleration zone.  Here \teq{\gamma = \{ 1+
(p/mc)^2 \}^{1/2} \approx p/(mc)} for energetic leptons. With
\teq{\gamma_{\rm max}} determined as the smaller of the limiting values
from the two constraints above, the high-energy electron distribution
{\it injected} at the acceleration site is written as \teq{n_{\rm acc}
(\gamma) \propto \gamma^{-\sigma (p )} \, \exp\left(-\gamma /
\gamma_{\rm max} \right)} for \teq{\gamma\gg 1}, where \teq{\sigma} is
the high-energy index of the {\it simulated} Monte Carlo distribution
\teq{dn/dp \propto p^{-\sigma}} of accelerated particles.

The radiation module then assumes an equilibrium between the
acceleration of particles out to \teq{\gamma_{\rm max}} and particle
escape out of the radiation zone on a time scale \teq{t_{\rm esc} =
\eta_{\rm esc}\, R/c}. If \teq{t_{\rm rad} <  t_{\rm esc}}, a break in
the electron spectra by \teq{\Delta\sigma = 1} is expected at a Lorentz
factor \teq{\gamma_b<\gamma_{\rm max}}, where the radiative cooling time
scale equals the escape time scale. This radiative cooling break is in
addition to acceleration flattening breaks illustrated in
Fig.~\ref{fig:accel_dist}. The resulting electron spectra are then
re-normalized to a kinetic luminosity of electrons in the jet,
\begin{equation}
   L_{\rm kin} \; =\; \pi R^2 c \, \Gamma^2_{\rm jet} \, m_e c^2 
   \int_{1}^{\infty} n_e (\gamma) \, \gamma \, d\gamma
 \label{eq:Lkin}
\end{equation}
which is a free input parameter for fitting purposes.  Here
\teq{\Gamma_{\rm jet}} is the bulk Lorentz factor of the jet.  As a
change in the electron spectrum (and density) will lead to changes in
the co-moving synchrotron photon field, our code determines the
equilibrium electron distribution through an iterative process.  It
starts by considering only the magnetic \teq{{\cal U}_B} and external
radiation \teq{{\cal U}_{\rm rad}} energy densities to determine a
first-order equilibrium \teq{e^-} distribution. Then \teq{{\cal U}_B} is
used to evaluate the co-moving synchrotron photon field, which is added
to the energy densities to re-determine the electron cooling rates and
to re-evaluate the break and maximum electron energies. The process is
repeated until convergence is achieved, in just a few iterations.

The magnetic field is specified by means of a magnetic partition
fraction $\epsilon_B \equiv L_B/L_{\rm kin}$, where \teq{L_B = \pi R^2
\, c \, \Gamma_{\rm jet}^2 \, {\cal U}_B} is the power carried by the
magnetic field along the jet, partly in the form of Poynting Flux. The
field is assumed to be tangled in the co-moving jet frame, and the
radiative output from the equilibrium electron distribution is evaluated
using the radiation transfer modules of Boettcher et al.\cite{Boettcher13}.

\subsection{Multiwavelength Spectral Model for AO 0235+164}
 \label{sec:spectra}

To illustrate the advances in understanding offered by combining shock
acceleration simulation results with radiation emission codes, we study
here the low synchrotron peak frequency BL Lac object AO 0235+164 at
\teq{z=0.94}. It belongs to a group of extragalactic jet sources with
the \teq{\nu F_{\nu}} spectrum peaking in the optical.  However,
interestingly, in flaring, it is observed to possess a significant X-ray
excess, consistent with the so-called {\it Big Blue Bump} (BBB).  Such
enhancements have been suggested\cite{SMMP97} to be a signature of a
bulk Comptonization effect, where hot thermal electrons upscatter an
ambient quasi-thermal radiation field, perhaps in the infra-red or
optical bands. This seed field could be disk-related, or perhaps dust
emission possibly from the line regions, but is distinct from the
non-thermal synchrotron emission.  Here, one of the objectives is to
explore whether the BBB can be attributed to the substantial thermal
electron pool evinced in shock acceleration predictions such as in
Fig.~\ref{fig:accel_dist}.  In the fast-moving jet, this population may
produce significant radiative signatures due to Comptonization of an
external radiation field.

\begin{figure}[pt]
\centerline{\includegraphics[width=12.0cm]{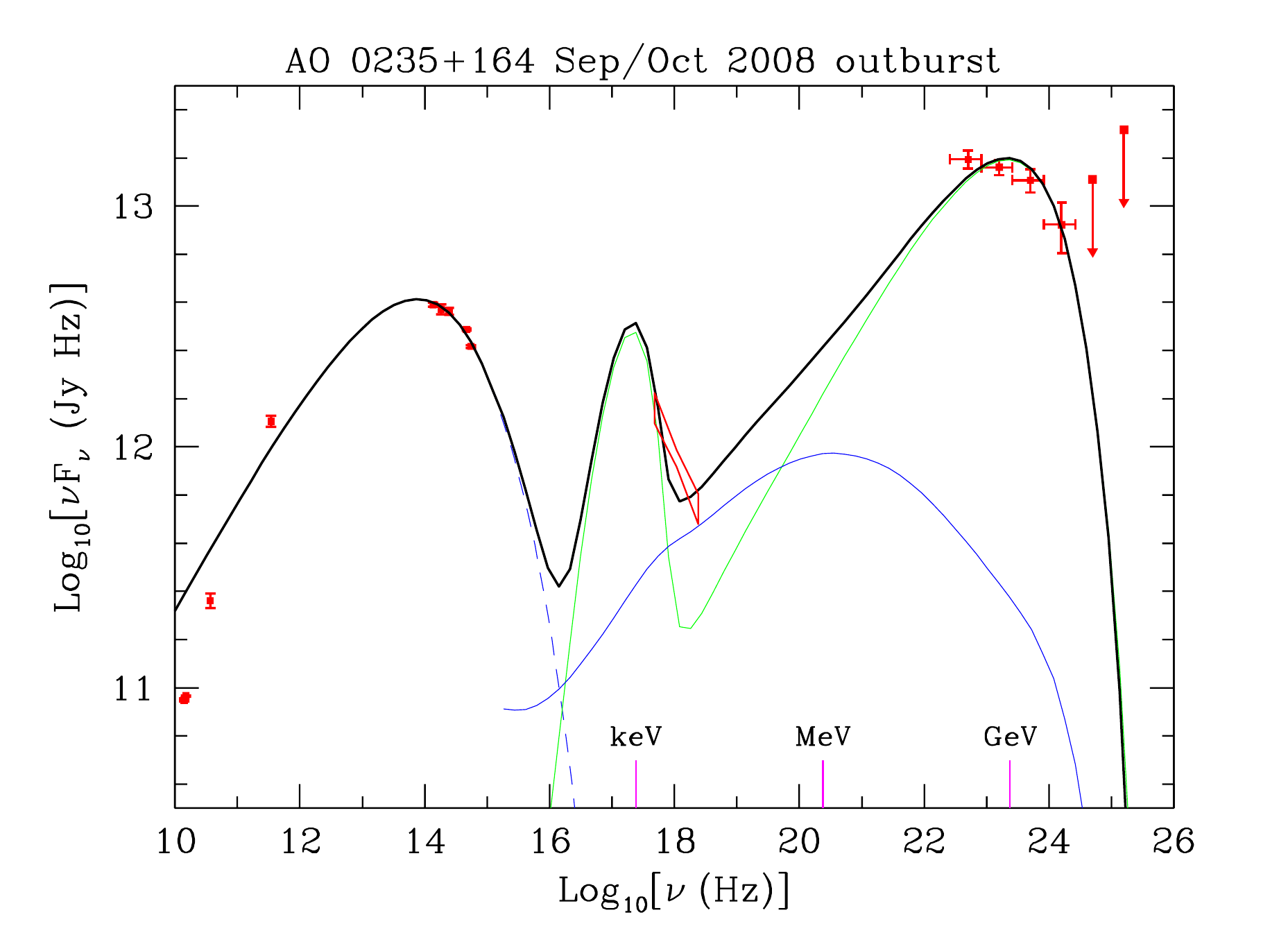}}
\vspace*{-5pt}
\caption{Multiwavelength \teq{\nu F_{\nu}} spectra (points) spanning the radio, 
optical, X-ray and gamma-ray bands, together with model fits as described in 
the text, for the September/October 2008 flaring event of the blazar AO 0235+164.  
The campaign data are taken from Ackermann et al.$^{41}$, displaying
a high state epoch for October 22-24.  The gamma-ray detections and upper limits 
are from {\it Fermi}-LAT, while the X-ray ``butterfly'' block represents {\it Swift} XRT data. 
The broadband models consist of a synchrotron component (dashed blue curve) up to
the optical band, an SSC component in the X-rays and gamma-rays (solid blue curve),
bulk Comptonization emission (green solid curve), totaling the black spectrum.  
The data and spectral model constrained the diffusion parameters to 
\teq{\eta_1=15} and \teq{\alpha =3}.
\label{fig:MW_spec_AO_0235}}
\end{figure}

The BL Lac object AO 0235+164 was observed to be flaring in the period
during September -- November 2008. We focus here on modeling data
collected during the multi wavelength campaign discussed in Ackermann et
al.\cite{Ackermann12}; see also Agudo et al.\cite{Agudo11} for
additional flare light curves and polarization data. The list of
observatories and missions participating in this campaign is
extensive\cite{Ackermann12} --- it spans radio and optical, with {\it
Swift} XRT and RXTE providing X-ray monitoring, and {\it Fermi}-LAT
providing the principal gamma-ray observations.  The LAT gamma-ray
spectrum displays a break or possible turnover at energies around
\teq{3-4}GeV, with a power-law index of about \teq{2.1} below this
feature; the source is not detected in the TeV band\cite{Errando11}. The
broadband spectrum for the high state portion of the outburst, during
the interval MJD 54761-54763 (October 22-24, 2008) is depicted in
Fig.~\ref{fig:MW_spec_AO_0235}.  This appears in Fig.~7 of Ackermann et
al.\cite{Ackermann12}, wherein it can be compared with a low state
spectrum obtained about six weeks later that does not contain a 
{\it Swift} XRT detection.
 
The jet parameters used to derive the model fit include the bulk Lorentz
factor of \teq{\Gamma_{\rm jet}=15}, and the electron kinetic luminosity
of \teq{L_{\rm kin}=1.55\times 10^{47}}erg/sec.  The magnetic luminosity
\teq{L_B} was a tenth of this so that the field was marginally
sub-equipartition: for this one-zone model, \teq{B\approx 0.45}Gauss was
obtained.  To place the synchrotron turnover frequency in the optical
band (\teq{\sim 10^{14}}Hz), the values of \teq{\eta (\pmax )} and
\teq{B} have been chosen so that the maximum electron Lorentz factor in
the comoving frame of the jet is \teq{\gamma_{\rm max}\approx 5.53\times
10^3}, a parameter connecting to the shock acceleration zone.  This then
placed the peak SSC frequency at around \teq{\gamma_{\rm max}^2
10^{14}\hbox{Hz}\sim 10^{21}}Hz, i.e. around 10 MeV, as is evident in
Fig.~\ref{fig:MW_spec_AO_0235}. A cooling break is apparent in the SSC
spectrum at around a few keV. Observe that the SSC component is of
insufficient luminosity to model the {\it Swift} XRT and {\it Fermi}-LAT
signals --- another component is needed to explain them.

Here we model these detections using bulk-Comptonization/inverse Compton
scattering of an external radiation field.  This field is presumed to be
quasi-isotropic in the observer's frame, so that the jet material moves
at \teq{\Gamma_{\rm jet}=15} relative to these seed photons, and so can
inverse Compton scatter them, i.e. increase their frequency by a factor
of \teq{\sim 4 \Gamma_{\rm jet}^2}.  The minimum Lorentz factor of an
electron accelerated in the distributions of Fig.~\ref{fig:accel_dist}
is less than 2, so this adds at most a factor of two to this IC energy
enhancement.  Hence, to explain the {\it Swift} XRT flaring flux, the
seed background radiation needed a temperature of \teq{T\approx 6\times
10^3}K, with a Planck energy density of \teq{{\cal U}_{\rm
rad}=1.1\times 10^{-5}}erg cm$^{-3}$. This corresponds to optical
radiation, perhaps from the accretion disk.  The thermal portion of the
electron population boosts the ambient photons up to the X-ray range,
and roughly produces the steep {\it Swift} XRT spectrum if it
corresponds to the upper side of the Maxwell-Boltzmann distribution. 
This is clearly evident in Fig.~\ref{fig:MW_spec_AO_0235}.  This then
connects to a very flat external Compton tail, generated by the broken
power-law tail portions of the distributions depicted in
Fig.~\ref{fig:accel_dist}, modulo the cooling breaks discussed above. 
This IC  tail possesses a spectral index of around 3/2, rising in the
\teq{\nu F_{\nu}} representation to meet the {\it Fermi}-LAT flux above
100 MeV.  The details of the various spectral components are outlined at
greater length in B\"ottcher et al.\cite{BBS12}. The key
acceleration/diffusion parameters for the fit will be discussed below.
To leading order, this one-zone hybrid acceleration/radiation model
describes the multi wavelength spectrum of AO 0235+164 in its high
outburst state, but only using a contribution of external-radiation
Comptonization from the jet material, a conclusion similar to that in
Ackermann et al.\cite{Ackermann12}.

\newpage

\section{Discussion}
 \label{sec:discussion}
 
The multiwavelength spectral fits obtained here using full electron distributions 
derived from diffusive shock acceleration simulations enable useful diagnostics 
on the plasma environment in blazar jets.  The key constraints derived are 
representative values of \teq{\eta (p) = \lambda_{\parallel}/r_g}
at the low injection momenta, \teq{p\sim p_1}, and at the maximum 
momentum \teq{\pmax} in the shock acceleration zone.  These two 
mean free path parameters couple via the diffusion index \teq{\alpha}.
Here we focus first on \teq{\eta (\pmax )}, which for our blazar case study
far exceeds unity.  The reason for this is that it is required to fit the frequency 
of the synchrotron \teq{\nu F_{\nu}} peak just below the spectral turnover of this 
emission component. For strong cooling in the acceleration zone, the turnover is 
created when the cooling rate \teq{-d\gamma/dt = 4 \sigt \, {\cal U} \, \gamma^2/(3 m_e c)}
is approximately equal to the particle acceleration rate 
\teq{d\gamma /dt\vert_{\rm acc} \sim (u_{1x}/c)^2\, eB/(\eta\, mc)} 
in the Fermi process.  Now introduce a cooling parameter \teq{\epsilon_{\rm syn}
= {\cal U}_B/{\cal U}} that represents the fractional contribution of the synchrotron process 
to the electron cooling.  This is approximately
\teq{\epsilon_{\rm syn}\approx L_O/L_{\gamma}}, which has a value of
about \teq{0.12-0.15} according to the broadband spectrum in Fig.~\ref{fig:MW_spec_AO_0235}.
If \teq{\epsilon_{\rm syn}\gtrsim 1}, then this acceleration/cooling equilibrium 
establishes the well-known result that \teq{\gammax \propto B^{-1/2}} and 
the synchrotron cutoff frequency is independent of the magnetic field strength:
\begin{equation}
  E_{\rm syn} \;\sim\; \frac{\Gamma}{\eta}
     \left( \dover{u_{1x}}{c} \right)^2 \, \dover{m_ec^2}{\fsc}\quad 
 \label{eq:Esyn_max}
\end{equation}
Here \teq{\fsc =e^2/(\hbar c)}, and the blueshift factor
\teq{\Gamma\equiv \Gamma_{\rm jet}} due to Doppler beaming has been
included. This result holds provided that the acceleration process is
gyroresonant, which is the prevailing paradigm for both non-relativistic
and relativistic shocks.  For \teq{\Gamma =1} and \teq{u_{1x}=c}, the
\teq{\eta (\pmax )=1} Bohm limit of this is around 100 MeV, as was
highlighted in De Jager et al.\cite{DeJager96} for considerations of
$\gamma$-ray emission at relativistic pulsar wind nebular shocks.  For 
\teq{\epsilon_{\rm syn} < 1}, as is the case for AO 0235+164,
\teq{\pmax} becomes weakly dependent on \teq{B}, and the turnover energy
\teq{E_{\rm syn}} drops somewhat below the estimate in
Eq.~(\ref{eq:Esyn_max}), and becomes proportional to \teq{B}.

For our application to AO 0235+164, we set \teq{\Gamma\equiv \Gamma_{\rm
jet} =15} and \teq{u_{1x}/c = 0.71}.  To move \teq{E_{\rm syn}} into the
X-ray or optical bands, one requires very large values for \teq{\eta
(\pmax )}, thereby dramatically reducing the rapidity of the
acceleration process. This was the approach of Inoue \&
Takahara\cite{IT96}, who chose \teq{\eta\sim 10^5} when exploring
multiwavelength modeling of Mrk 421 spectra. For the AO 0235+164 fitting
protocol here, one needs much larger values (\teq{\eta\sim 10^8}) to
place \teq{E_{\rm syn}} in the optical band. Values this large suppress
injection of thermal charges into the acceleration process almost
completely\cite{SB12}.  Therefore the path to maintaining efficient
injection but slowing acceleration out to the maximum momenta is to
demand that \teq{\eta} not be constant, but rather a strongly increasing
function of \teq{\gamma}. This was adopted in the fits of
Fig.~\ref{fig:MW_spec_AO_0235}, where \teq{\eta_1=15} and \teq{\alpha=3}
(i.e. \teq{\eta (p) \propto p^2}) yields a synchrotron turnover in the
optical.  Such constraints on the shock turbulence parameters provide a
strong Maxwell Boltzmann component to the electron distribution, as in
Fig.~\ref{fig:accel_dist}, and this generates a bulk Comptonization peak
in the X-ray band that accommodates the {\it Swift} XRT spectrum and
flux.  At the same time, since \teq{\eta (p)\gg 10} for most electron
momenta, the distribution tail quickly asymptotes to a power law with an
index \teq{\sigma (p)\sim 1}, corresponding to very active shock drift
acceleration in weakly turbulent fields, as discussed above. This then
generates the \teq{3/2} photon index evident in the non-thermal external
Compton component, when the population is subjected to strong Compton
and synchrotron cooling outside the acceleration zone.  Since AO
0235+164 does not have a TeV detection, its inverse Compton peak energy
is low enough to preclude a determination of \teq{\sigma (\pmax )} using
the {\it Fermi}-LAT instrument.

The multi-wavelength fits indicate that the physical environment of the
shock acceleration region should possess weakening turbulence at larger
distances from the discontinuity, driving longer diffusive mean free
paths for electrons of larger momenta.  This is not entirely without
precedent.  In the heliosphere, the solar wind contains traveling
interplanetary shocks, at which accelerated ions are clearly detected. 
The turbulence near the shocks is observed to satisfy \teq{\langle\delta
B/B\rangle \sim 0.1}, and the diffusive transport of H and He ions is
inferred to be not very different from the Bohm limit\cite{BOEF97}.  In
contrast, away from these shocks, the turbulence is much more muted, and
the diffusive mean free paths are much larger\cite{FJO74,Palmer82}. 
From this, \teq{\alpha >1} circumstances must clearly exist.  Actual
magnetometer measurements\cite{PRG07} by the {\it Wind} spacecraft of
the quiet solar wind over a period of several years establishes an
inertial range for turbulence \teq{\langle (\delta B)^2/8\pi \rangle}
spanning about 3-4 decades in frequency above \teq{\sim 3\times
10^{-5}}Hz. The spectrum for this range is fairly close to the standard
Kolmogorov \teq{\nu^{-5/3}} form.  Below this frequency, the turbulence
spectrum flattens, and one anticipates that diffusion in solar wind
plasma turbulence must decline.

From the theoretical standpoint, given a finite inertial (i.e. active)
range of turbulence, diffusive mean free paths \teq{\lambda_{\parallel}}
can be estimated\cite{FJO74} using quasi-linear theory (QLT).  For
Kolmogorov-like turbulence spectra, one can quickly arrive at
\teq{\lambda_{\parallel} \propto p^{2/3}}, approximately. However, for
charges with high momenta and therefore large gyroradii, the Doppler
resonance condition \teq{\omega \equiv kc = eB/(\gamma mc)} is not
satisfied by turbulence in the inertial range, only by low frequency
waves. Therefore, it can be simply demonstrated that the mean free paths
become more strongly dependent on \teq{p} using the formalism
of\cite{FJO74}, namely that approximately
\teq{\lambda_{\parallel}\propto p^2}, i.e. \teq{\alpha\sim 2} and
\teq{\eta \propto p}. The diffusion of charges then receives a
substantial contribution from non-gyroresonant collisions with inertial
turbulence.  Such is expected to be the situation for electrons with
large gyroradii being accelerated in blazar shocks.  Moreover, if the
shock-generated inertial range turbulence does not persist at the same
levels out to very large distances from blazar shocks, but instead
declines in intensity, one can anticipate that \teq{\alpha} can
effectively rise above two. In other words, \teq{\alpha >1} scenarios
are entirely realistic for blazar multi wavelength models.  Accordingly,
the broadband spectral modeling of AO 0235+164 presented here provides
interesting diagnostics on the nature of plasma turbulence in its
shocked jet.

\newpage

\section{Conclusion}

This paper has offered a detailed exploration of multi-wavelength
spectral fits to flaring emission from the BL Lac  object AO 0235+164
using electron distribution functions obtained directly from Monte Carlo
simulations of diffusive acceleration at relativistic shocks.  This
cohesive application of simulated acceleration distributions to blazar
spectral modeling provides essentially unprecedented insights into the
plasma environment of extragalactic jets.  Since shock acceleration
feeds off a prominent thermal population of charges, low synchrotron
``peak'' frequency sources like AO 0235+164 should possess strong bulk
Comptonization signatures from this thermal component.  Such is the case
here in Fig.~\ref{fig:MW_spec_AO_0235}, accounting for the high-state
{\it Swift} XRT signal.  At the same time, in order to model both the
low synchrotron turnover energy in the optical, and the prominent hard
gamma-ray flux seen by {\it Fermi}, models need the diffusive mean free
path to increase rapidly with electron momentum,
\teq{\lambda_{\parallel}\propto p^{\alpha}} with \teq{\alpha \gtrsim 2},
starting with fairly modest values \teq{\lambda_{\parallel}/r_g \sim
10-30} at low thermal momenta. We argue here that this strong momentum
dependence of \teq{\lambda_{\parallel}/r_g} is not unexpected, due to
somewhat constrained inertial ranges of plasma turbulence, associated
non-gyroresonant diffusion, and perhaps also a decline in the level of
such turbulence away from shocks that inject energetic particles into
the much larger blazar jet emission zones.

\section*{Acknowledgments}

MGB and MB are grateful to NASA for partial support for this research 
through the Astrophysics Theory Program, grant NNX10AC79G.
MGB also acknowledges support from the Department of
Energy under grant DE-SC0001481.
This work is based on research supported by the South African Research Chairs
Initiative (grant no. 64789) of the Department of Science and Technology and the
National Research Foundation of South Africa.

\vspace{20pt}


\end{document}